# Comparison of room-temperature multiferroics in $Bi_4Fe_2TiO_{12}$ film and bulk

J. Lu, L.J. Qiao and W.Y. Chu

*Department of Materials Physics, University of Science and Technology Beijing, Beijing 100083, People's Republic of China*

It is reported that both dielectricity and magnetism at room temperature have been appreciably improved in $Bi_4Fe_2TiO_{12}$ film compared with that of $Bi_4Fe_2TiO_{12}$ bulk. X-ray diffraction profiles reveal similar single-phase crystalline nature and random orientation of the two but X-ray photoelectron spectroscopy experiments indicate 1.4 eV lower binding energy of core-state O 1s in film relative to those of bulk, so the improvement of multiferroics in film are attributed to the oxygen vacancies and high fraction of interface. The results have promising applications in multifunctional integrated devices.

Multiferroics has become one of the time-honored issues in last years, herepolarization and spin order coexistent simultaneously in one entity [1-7].Considerable progresses in single-phase materials include direct manipulation of polarization reversal by magnetical field in single crystal $TbMn_2O_5$ [1], giant magnetodielectric in single-crystal $DyMnO_3$ [2], and colossal magnetodielectric effects in single-crystal $CdCr_2S_4$ [3]. However, so far, such strong multiferroic couplings have never been discovered in room temperature, although room-temperature strong coupling between magnetic field and polarization has recently reported in $LuFe_2O_4$ single crystal, where dielectric constant can transiently decreases by only 20% [4]. Epitaxial $FeBiO_3$ thin films have been synthesized with remarkable spontaneous polarization, magnetism and magnetoelectric output at room temperature [5]. The interaction between spontaneous polarization and external magnetic field in $Bi_{0.6}Tb_{0.3}La_{0.1}FeO_3$ single phase but thin film has also been suggested more pronounced with about 80% reduction of remanent polarization after magnetic-field treatment of 9 Tesla, as a consequence of magnetic-field-induced irreversible reorientation of grains [6]. However, the comparison between films and bulks in strong correlated multiferroic materials and uniqueness of multiferroics in film has rarely been discussed. This article presents room-temperature experimental results of enhanced dielectricity and magnetism in $Bi_4Fe_2TiO_{12}$ film in comparison with $Bi_4Fe_2TiO12$ bulk, wherein the origin of improvement is interpreted by the evidences from X-ray diffraction profiles and X-ray photoelectron spectroscopy (XPS).

$Bi_4Fe_2TiO_{12}$ films were spin-coating deposited onto the platinized silicon layer by layer from chemical solution, and single-phase polycrystalline film with a thickness of hundreds of nanometers was then obtained by rapid annealing. Before the electric measurement, which was carried out under the HP4194 dielectric spectra analyzer at room temperature, top electrodes were prepared via conventional mask technology. The measurement of magnetism is conducted in a superconducting quantum interference device (SQUID, by Quantum Design Inc.). The details of sample preparation have been published elsewhere [7].

To evaluate the solubility of Fe element at the pseudoperovskite B site in $Bi_4Ti_3O_{12}$, $Bi_4Fe_nTi_{3-n}O_{12}$ (n=0,1,2) film and powder derived from the corresponding sol was prepared and characterized via X-ray diffraction. Fig. 1 shows the results, revealing single-phase nature and no detectable signals of second phase for $Bi_4Fe_nTi_{3-n}O_{12}$ when n increases up to 2. The XRD profile of $Bi_4Ti_3O_{12}$ powder in Fig. 1f was indexed according to JCPDS file with ID 35-0795 and was found quite close to that of standard Bi-layered Aurivillius $Bi_4Ti_3O_{12}$ structure with B2cb space group [8]. So crystalline structure of $Bi_4Ti_3O_{12}$ film as well as powder is considered single-phase bismuth-layered Aurivillius structure, where Fe substitutes Ti uniformly into the bismuth layered lattice in its favorable way [9]. Additionally, by comparing the peak positions of $Bi_4Fe_2TiO_{12}$ film with those of $Bi_4Fe_2TiO_{12}$ ceramics prepared from $Bi_4Fe_2TiO_{12}$ powder, depicted in Fig. 2, one can find good correspondence between the two except for peaks from the substrate of the film, indicating single-phase nature of $Bi_4Fe_2TiO_{12}$ ceramics. It is understandable that $Bi_4Fe_2TiO_{12}$ polycrystalline film is random oriented since the intensity of corresponding peaks is the same order of magnitude with that of $Bi_4Fe_2TiO_{12}$ ceramics.

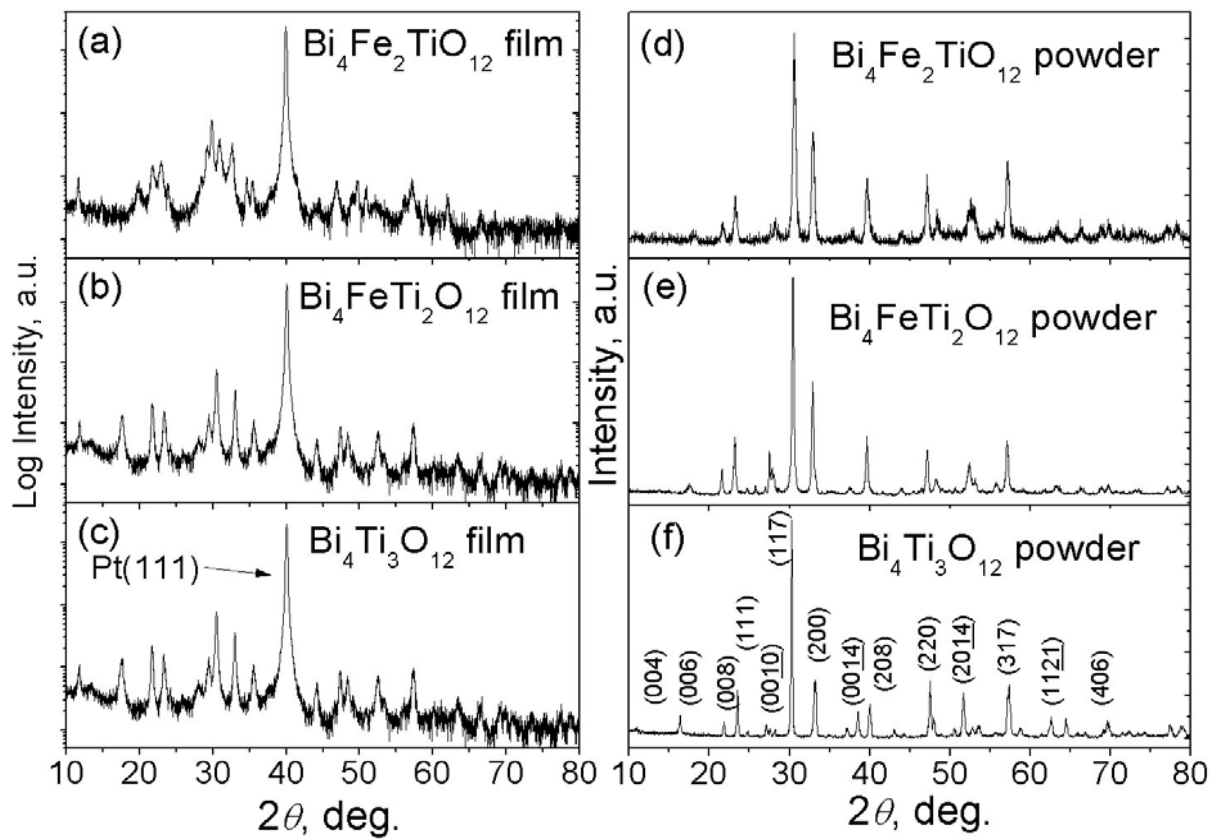

Fig. 1 X-ray diffraction profile of $Bi_4Fe_nTi_{3-n}O_{12}$ (n=0,1,2) film and corresponding powder derive from respective sol. The indexes in (f) are referred from JCPDS file with ID 35-0795.

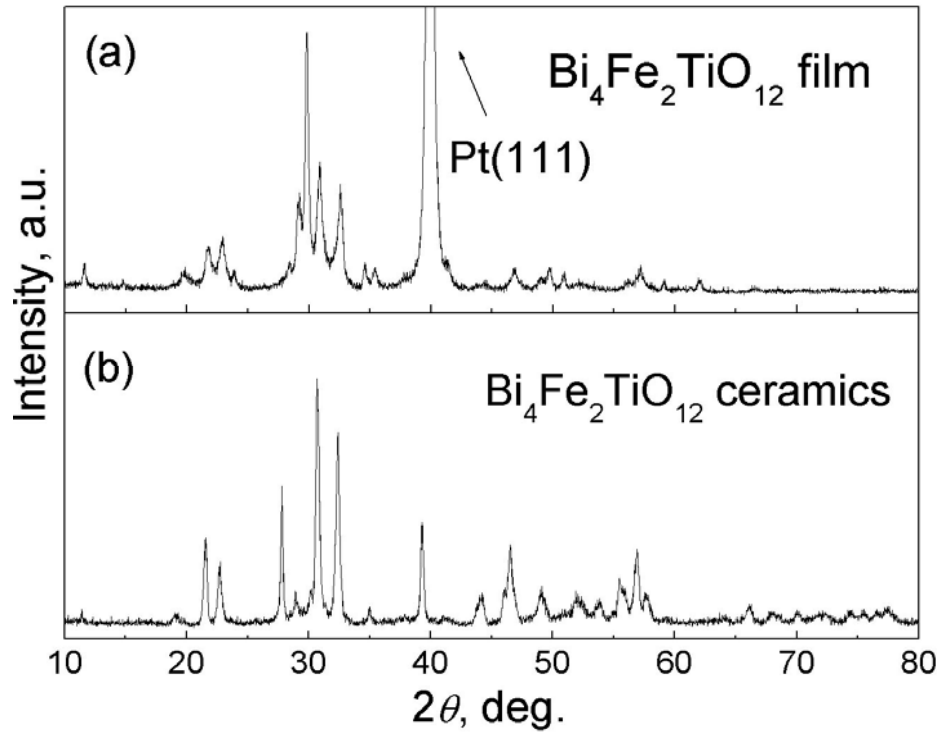

Fig. 2 X-ray diffraction profile of $Bi_4Fe_2TiO_{12}$ film and ceramic bulk

The results of functional measurement of $Bi_4Fe_2TiO_{12}$ ceramic bulk and film have been shown in Fig. 3 and Fig. 4. In Fig. 3, the dielectric constant of $Bi_4Fe_2TiO_{12}$ film at room temperature is approximately twice as much as that of $Bi_4Fe_2TiO_{12}$ ceramics in the frequency range of 100 Hz~100 kHz. Meanwhile, the magnetization of $Bi_4Fe_2TiO_{12}$ film is remarkably larger than that of $Bi_4Fe_2TiO_{12}$ ceramics up to room temperature, as shown in Fig. 4. The clear peak of $Bi_4Fe_2TiO_{12}$ film at about 50 K in Fig. 3 implies a transition from antiferromagnetism to weak ferromagnetism when warming, as indicated in another publication [7].

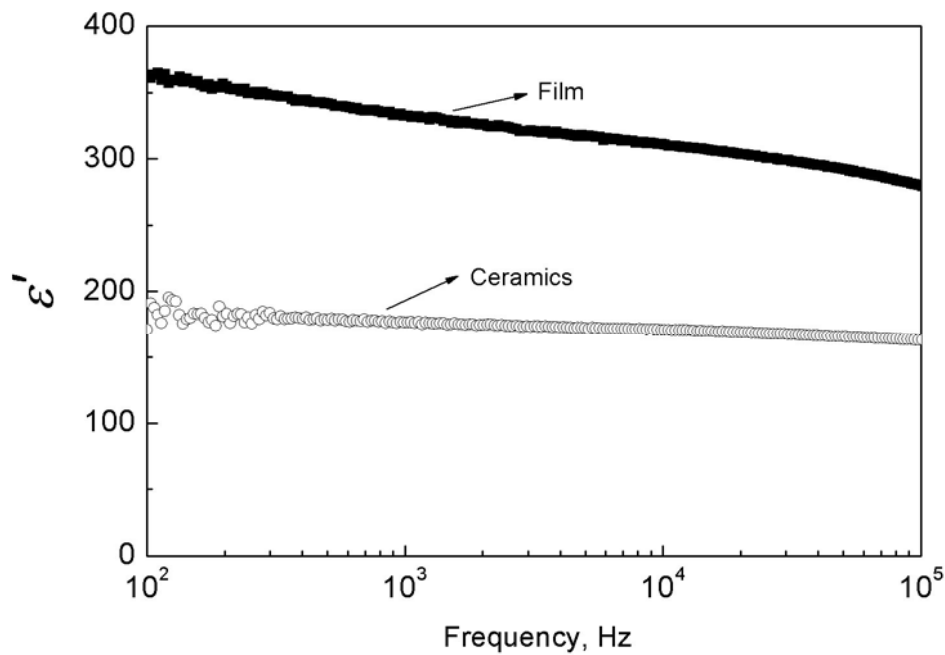

Fig. 3 Dielectric constant of $Bi_4Fe_2TiO_{12}$ film and ceramic bulk

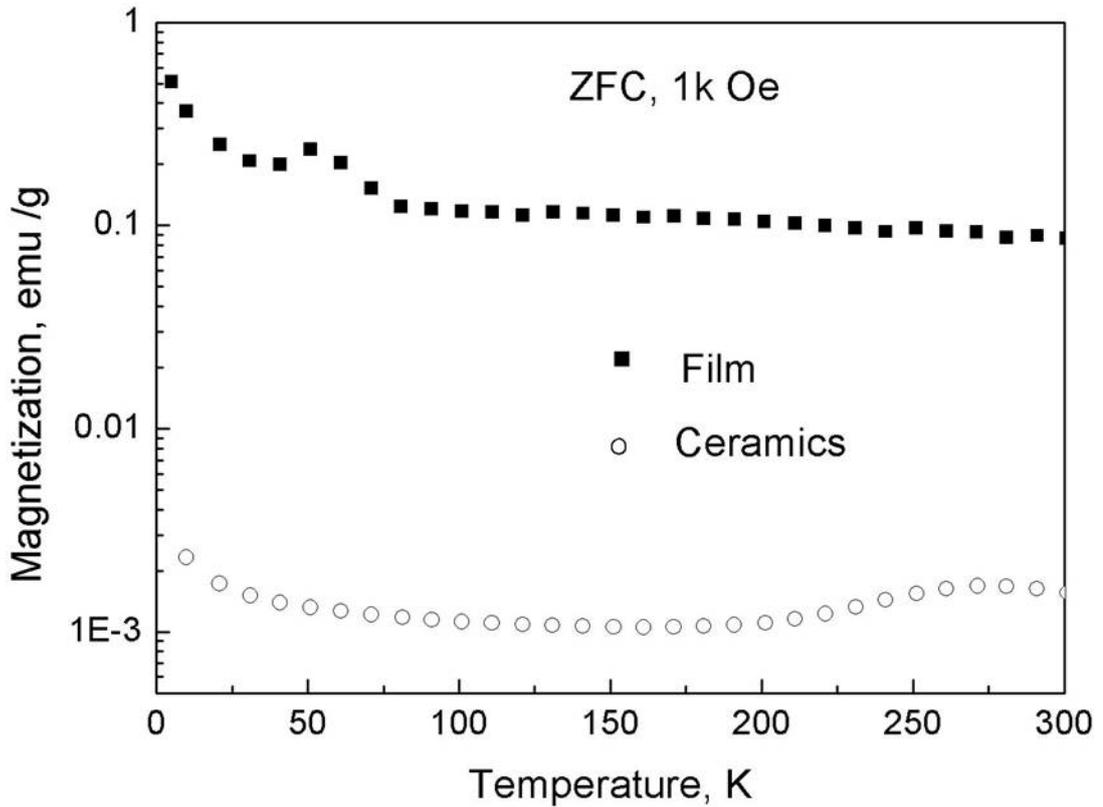

Fig. 4 Magnetization versus temperature curves of $Bi_4Fe_2TiO_{12}$ film and ceramic bulk

To analyze the XPS data, all of the binding energies at various peaks were calibrated by using the binding energy of C 1s (284.6 eV) [10], as shown in Fig. 5 and Fig. 6. In Fig. 5, the 7/2 and 5/2 spin-orbit doublet components of the Bi 4f core level photoemission are located at 158.9 and 164.2 eV, respectively, not only for $Bi_4Fe_2TiO_{12}$ film but also for $Bi_4Fe_2TiO_{12}$ ceramics when taking the error about 0.5 eV into account. These two peaks as well as the 5.3 eV spin-orbit splitting of Bi 4f core level are quite close to that of $Bi_4Ti_3O_{12}$ ceramics [11], suggesting the same chemical environment of $Bi_4Fe_2TiO_{12}$ film and ceramics to the $Bi_4Ti_3O_{12}$ ceramics and uniform substitution of Fe at Ti sites in $Bi_4Fe_2TiO_{12}$ film and ceramics. In contrast, the O 1s core level photoelectron energy of $Bi_4Fe_2TiO_{12}$ film is different from that of ceramics, as shown in Fig. 6, where the XPS peak of $Bi_4Fe_2TiO_{12}$ film are shifted to the lower-energy end by 1.4 eV in comparison with that of $Bi_4Fe_2TiO_{12}$ ceramics and furthermore a shoulder appears at about 532 eV on the higher energy side of $Bi_4Fe_2TiO_{12}$ film. The two major features in $Bi_4Fe_2TiO_{12}$ film have been clearly shown by Gaussian peak fitting where the sum of the two sub fitting peaks agrees quite well with the experimental results. Intuitively, the decrease of O 1s core-level binding energy is attributable to the screening effect, where emitted photoelectrons leave behind a positive-charged hole which is screened by relaxation of surrounding atoms [12]. Strictly speaking, in the view of band theory, the promotion of oxygen vacancies, which act as donors in oxide, causes an increase in surface conductivity and a band bending [13]. Another feature of XPS plot of presence of a side peak at about 532 eV in $Bi_4Fe_2TiO_{12}$ film should not be due to hydroxyl groups, or to any other impurities, but arises from the effect of oxygen vacancies [14]. The assignment of this O 1s shoulder at 532 eV to O2- anions located near to oxygen vacancy sites in film has been also claimed by another separated group [15].

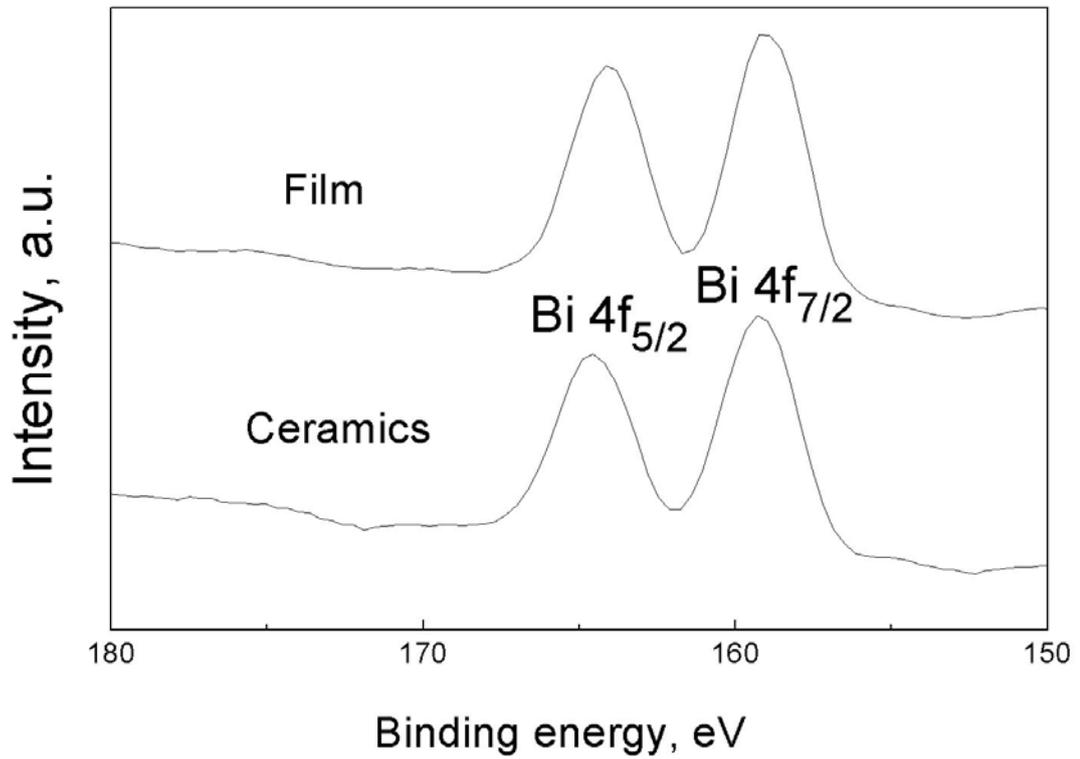

Fig. 5 Core level photoelectron spectra of Bi 4f of $Bi_4Fe_2TiO_{12}$ film and ceramic bulk

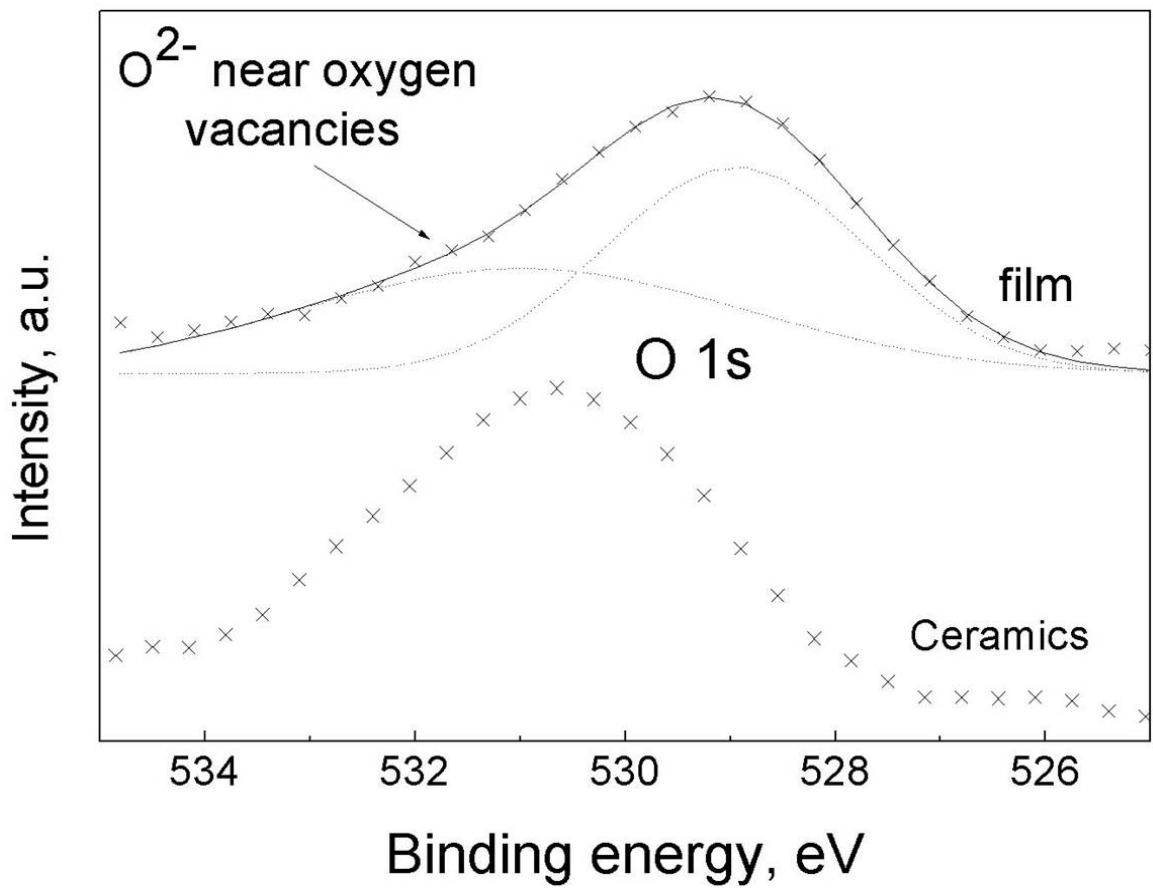

Fig. 6 Core level photoelectron spectra of O 1s of $Bi_4Fe_2TiO_{12}$ film and ceramic bulk. Crosses indicate experimental data solid and dashed curves show the sum and sub Gaussian fitting resuls, respectively.

It is not unusual for the occurrence of a higher dielectric constant in perovskite oxide film when compared with ceramics because of various factors such as stresses from the substrate and dielectric inhomogeneities like interface layers, grain boundaries and porosity [16]. Actually, in bismuth layered materials, rotation of defect dipoles associating the oxygen vacancies toward the direction of applied electric field causes the majority of polarizability [17]. Hence, the increase of dielectric constant in Fig. 3 can be explained by the more oxygen vacancies in $Bi_4Fe_2TiO_{12}$ film, which is consistent with Fig. 6.

Intriguing, the room striking improvement of magnetism in $Bi_4Fe_2TiO_{12}$ film, which is hardly due to the minor impurity or orientation effect in film as shown in Fig. 4, can be understood by the unique room-temperature ferromagnetic phenomena in film. J.M.D. Coey et al. advocated that room-temperature ferromagnetism of film arises from shallow donor electrons that form bound magnetic polarons, which is vacancies mediated and overlap to create a spin-split impurity band [18]. Therefore, we propose that the nontrivial room-temperature ferromagnetism $Bi_4Fe_2TiO_{12}$ film is also attributable to the number of oxygen vacancies, which is consistent not only with dielectric characterization in Fig. 3, but also with XPS results in Fig. 6.

Conclusively, we have synthesized $Bi_4Fe_2TiO_{12}$ films with appreciably promoted room-temperaturedielectric constant and magnetism in comparison with $Bi_4Fe_2TiO_{12}$ bulk via a facile chemical deposition method. The presence and interaction of oxygen vacancies is speculated a critical feature in $Bi_4Fe_2TiO_{12}$ films to produce enhanced correlated properties. The reported materials and phenomena are promisingly applicable in integrated multifunctional devices.